# Supercoherence: Harnessing Long-Range Interactions to Preserve Collective Coherence in Disordered Systems

Alexey Gorlach[1], Andrea Pizzi[2], Klaus Mølmer[3], Joseph Avron[1], Mordechai Segev[1] and Ido Kaminer[1]

[1]*Solid State Institute, Technion – Israel Institute of Technology, Haifa 32000, Israel*
[2]*Cavendish Laboratory, University of Cambridge, Cambridge CB3 0HE, United Kingdom*
[3]*Niels Bohr Institute, The University of Copenhagen,* 2200 *Copenhagen, Danmark*

**Artificial quantum systems with synthetic dimensions enable exploring novel quantum phenomena difficult to create in conventional materials. These synthetic degrees of freedom increase the system's dimensionality without altering its physical structure, accessing higher-dimensional physics in lower-dimensional setups. However, synthetic quantum systems often suffer from intrinsic disorder, causing rapid decoherence that limits scalability—a major obstacle in quantum information science. Here, we show that introducing just a few long-range interactions can mitigate decoherence, creating persistent collective coherence in highly symmetric collective excited states. We term this universal phenomenon "supercoherence" and show its exceptional robustness against disorder up to a dynamical phase transition at critical interaction strength and disorder. Supercoherence stabilizes not only coherence but also all other quantum properties of the states, challenging traditional views on the inevitability of decoherence in disordered interacting quantum systems and suggesting new opportunities for quantum memory and information processing.**

The concept of synthetic dimensions has emerged as a powerful tool for creating artificial systems with novel quantum states[1–3]. This approach increases a system's dimensionality without altering its physical structure, which can be used to create exotic topological phenomena[4–7]. Synthetic dimensions can be implemented in various platforms, including optical[6–13], atomic[4,14–17] and molecular[18] systems, and can be encoded on center-of-mass degrees of freedom, like the transverse or angular momentum of photons and atoms[18], or internal degrees of freedom, such as atomics hyperfine[14,15] and electronic[16] states. Synthetic dimensions can also be created in Floquet systems utilizing time-periodic modulations[19–23].

Synthetic dimensions help realize complex quantum states that are impossible to create in conventional systems with one, two, or three dimensions. Synthetic systems can utilize long-range interactions to engineer effective *infinite* dimensionality[24]. These interactions can be realized in various experimental platforms, such as atomic ensembles and arrays interacting through optical cavities and waveguides[25–28], color centers and quantum dots interacting through various photonic structures[29–31], superconducting qubits interacting through microwave waveguides[32], and trapped ions interacting through phonon waves[33]. The ability to control the range and strength of the interactions has led to the observation of fascinating phenomena, from inducing superconductivity in otherwise insulating materials[34] to quantum droplets and supersolid phases in ultracold atomic gases[35]. There is great interest in finding novel states and phases of matter facilitated by long-range interactions also in synthetic systems.

Here we show that synthetic systems with long-range interactions can undergo a dynamical phase transition[36,37] into a phase in which a particular set of excited states maintain their collective coherence under disorder, a phenomenon we term "supercoherence". We show that these states of robust coherence are manifested in a wide range of quantum systems with different interaction geometries, always exhibiting the same universal behaviors. In the special case of all-to-all interactions, supercoherence is closely related to dynamical phase transitions in superconductivity and cavity quantum electrodynamics[38]. We show that even sparse network geometries with a few long-ranged interactions can support supercoherence. We analyze the robustness of the supercoherent states and find the critical ratio of interaction strength to the level of disorder, underpinning the dynamical phase transition. Finally, we show that supercoherent states can maintain quantum memory in macroscopic superpositions, due to the robustness of both individual states and their superpositions.

**Collective quantum coherence and its implications**

Collective coherence in ensembles of quantum particles plays a crucial role in chemistry and physics[39,40]. Classical coherence underlies the sensitivity in magnetic resonance imaging and spectroscopy, with applications in medical diagnostics and chemical analysis[41]. Phenomena such as superradiance and superfluorescence[42–46] are often attributed to collective coherence in quantum dots, superlattices and molecular aggregates[47–49]. Other manifestations of collective coherence include time crystals[20,21] and superabsorbtion[50]. In atomic physics, collective quantum coherence is critical for metrology and sensing[51,52], while related concepts in quantum dots and vacancy centers suggest long-lived quantum memory[53,54].

The preservation of coherence is, however, challenged by decoherence. Its most common cause is inhomogeneous broadening, i.e., disorder in the individual transition frequencies, which is influenced by interparticle interactions, temperature fluctuations, variations in local screening, and size inhomogeneities. In synthetic systems like quantum dots and superconducting qubits, a large frequency variance limits applications in quantum information science and technology.

A significant contribution toward the extension of coherence in atomic systems, known as "cavity protection", was proposed theoretically[52–55] and demonstrated experimentally[31], showing how decoherence induced by inhomogeneous broadening can be suppressed. Nevertheless, in *general* quantum systems, reducing disorder is still a frontier challenge in many material platforms and especially in synthetic systems.

Our work proposes a path for robust collective coherence in the presence of disorder, by exploiting long-range interactions. We show that even sparse connectivity (e.g., a few interactions per particle) can be sufficient to induce a dynamical phase transition into a supercoherence phase. In this phase, specific highly excited states of systems with long-range interactions exhibit collective coherence that no longer decays due to inhomogeneous broadening. The collective coherence of such robust states is only limited at longer timescales by inevitable coupling to other decoherence channels, e.g., spontaneous emission of photons or phonons.

To analyze this phenomenon quantitatively, we present a model of two-level systems (spins) with frequency disorder and interactions (Fig. 1). This model and its close variants qualitatively apply to superconducting qubits, clusters of quantum dots, arrays of neutral atoms, and ions, as well as chemical aggregates and arrangements of biomolecules.

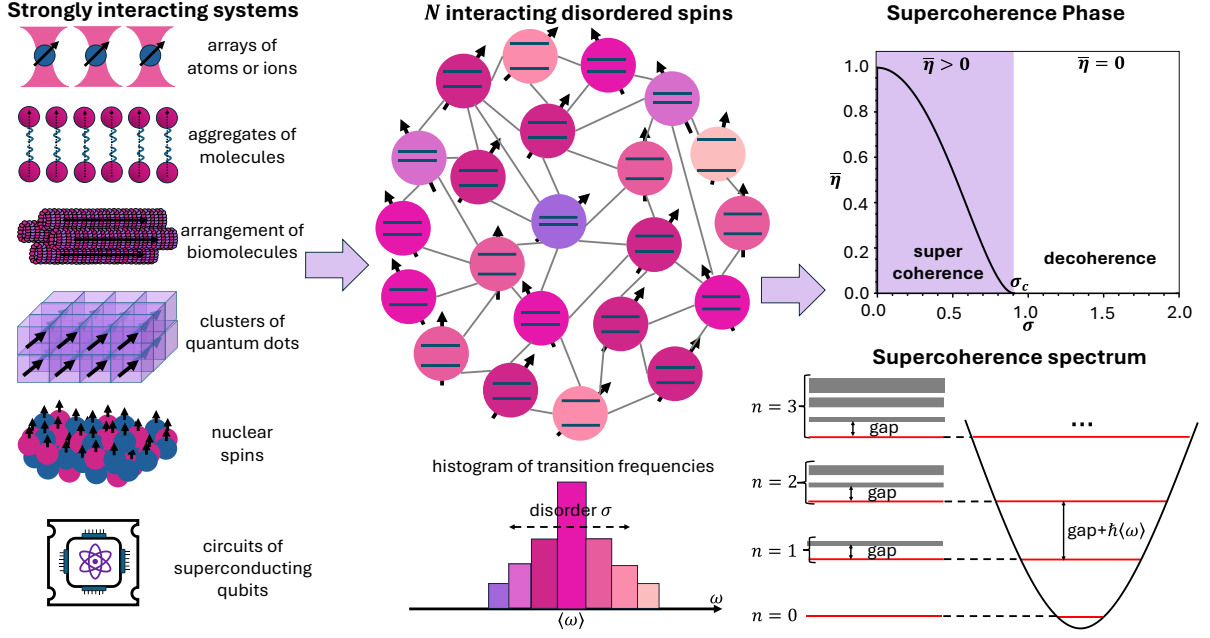

**Fig. 1: Disordered ensemble of two-level spins**. The figure illustrates an ensemble of interacting two-level spins, characterized by the frequency disorder in their transition frequencies and their interaction strengths. This model applies to various physical scenarios including clusters of quantum dots, superconducting qubits, aggregates of molecules, and arrays of neutral atoms. For long-range interactions, the system exhibits supercoherence, that is, an infinitely long-lived collective coherence $\bar{\eta}$. The supercoherence phase ($\bar{\eta} > 0$) and decoherence phase ($\bar{\eta} = 0$) are separated by a dynamical phase transition at a critical disorder $\sigma_c$. The supercoherent system has an energy spectrum with multiple bands and isolated delocalized states between them. These isolated states approximate the spectrum of a quantum harmonic oscillator, which remains stable against disorder and enables storing complex quantum states.

### The model

We consider $N$ spins with randomly distributed transition frequencies $\omega_i$ and hopping interactions $\tilde{J}_{ij} = \tilde{J}_{ji}$ for $i,j = 1,2,\ldots N$. The Hamiltonian reads $H = \hbar \sum_i \omega_i s_i^z + \hbar \sum_{i \neq j} \tilde{J}_{ij} s_i^+ s_j^-$, with $s_i^z$ and $s_i^\pm = s_i^x \pm s_i^y$ the spin 1/2 (Pauli) operators. We introduce the average interaction strength $J = (N-1)^{-1} \sum_{i \neq j} |\tilde{J}_{ij}|$, move to a rotating frame, and write the Hamiltonian in units of $\hbar J$, obtaining:

$$H = \sum_i \Omega_i s_i^z + \sum_{i \neq j} J_{ij} s_i^+ s_j^-, \qquad (1)$$

where $\Omega_i = (\omega_i - \langle \omega \rangle)/J$, $\sum_i \Omega_i = 0$, $J_{ij} = \tilde{J}_{ij}/J$, $\sum_{i \neq j} |J_{ij}| = N-1$, and $\langle \omega \rangle = N^{-1} \sum_i \omega_i$. Models like Eq. (1) are common in condensed matter physics, e.g., the XY-model, used to study many-body localization[56], usually considering lattice geometries and short-range interactions. The model captures decoherence by frequency disorder via a distribution of $\Omega_i$. Other external decoherence channels and losses are assumed here to occur on a longer timescale but can be incorporated using a Lindblad formalism as shown in Supplementary Materials (SM) Section 2.

We define the collective coherence order parameter as the average $xy$-component of the total spin:

$$\eta(t) = \frac{4}{N^2} \text{Tr}[S^+ S^- \rho(t)], \qquad (2)$$

where $S^\pm = \sum_i s_i^\pm$ and $\rho(t)$ is the density matrix describing the system. A mean (time-independent) order parameter can be defined upon time-averaging, $\bar{\eta} = \lim_{T \to \infty} T^{-1} \int_0^T \eta(t) \, dt$[38]. The collective coherence $\eta(t)$ describes the amplitude of the total spin projection on the $xy$-plane, a natural measure of collective coherence[57], which is directly measurable, for example as the magnetization in nuclear magnetic resonance spectroscopy. For large systems ($N \gg 1$) and no interactions ($J_{ij} = 0$), any frequency disorder $\Omega_i$ leads to a decay of $\eta(t)$ to zero (see SM, Section 1.2).

The interactions $J_{ij}$ can halt the decay of coherence despite the disorder due to the existence of a unique set of supercoherent states $|n_{SC}\rangle$, each separated by an energy gap from a band containing the other states. These isolated supercoherent states closely approximate the symmetric states $|n_s\rangle = (\text{norm})^{-1/2}(S^+)^n|0\rangle$, with $|0\rangle = \prod_i |0\rangle_i$ being the ground state of all spins. The symmetric states are widely used in studies of superradiance, superfluorescence, and other effects in condensed matter physics [42–46]. Each such state is simultaneously an eigenstate of $S_z$ and $S^+S^-$, giving the largest collective coherence according to Eq. (2).

Below, we explore the supercoherence phenomenon from two different directions: (I) We use mean-field theory to examine the transition from the decoherence phase ($\bar{\eta} = 0$) to the supercoherence phase ($\bar{\eta} > 0$). We show that supercoherence exists for different probability distributions of frequencies $\Omega_i$ and different initial conditions, emphasizing the universality of this phenomenon. We further analyze what initial spin states can lead to supercoherence and demonstrate its existence beyond mean-field theory. (II) We then investigate the full quantum description of supercoherence and its emergence in a wide range of interaction geometries. For each interaction geometry, we conduct a Hamiltonian spectral analysis of the energy levels and gaps by solving the system in the low-excitation regime. These complementary approaches help us identify the general properties of the supercoherence phenomenon and the conditions for its emergence, in both semi-classical and fully quantum models across various interaction geometries.

**Supercoherence in mean-field theory and beyond**

Our first analysis of supercoherence uses the mean-field approximation, for which we consider the limit $N \to \infty$ with all-to-all interactions $J_{ij} = -N^{-1}$. Physically, this extreme case has each spin undergoing precession with two components: a rotation around the z-axis with a random frequency $\Omega_i$, and a rotation around a perpendicular axis determined by the mean field generated by all other spins. In this way, we neglect quantum fluctuations, that are small for large $N$. In SM Section 3 we show that the results of this section hold also beyond the mean-field approximation. Below a certain frequency disorder $\sigma_c$, we identify a transition from the *decoherence phase* ($\bar{\eta} = 0$) to the *supercoherence phase* ($\bar{\eta} > 0$). This dynamical phase transition is marked by a jump in the derivative of the order parameter $\bar{\eta}$, akin to conventional phase transitions.

As an initial condition, we consider a separable symmetric state, with the same density matrix for each spin $\rho(r_0, \theta_0, \phi_0) = 1/2 + \vec{r} \cdot \vec{s}_i$, where $\vec{s}_i$ is the vector of spin matrices, $\vec{r} = (r_0 \sin\theta_0 \cos\phi_0, r_0 \sin\theta_0 \sin\phi_0, -r_0 \cos\theta_0)$ and $r_0 \in [0,1]$ quantifies the spins' purity. For $N \to \infty$, the spins always remain in a product state $\rho(t) = \prod_i (1/2 + \vec{r}_i(t) \cdot \vec{s}_i)$ and conserve the purity $r_0$, where $\vec{r}_i = (r_0 \sin\theta_i(t) \cos\phi_i(t), r_0 \sin\theta_i(t) \sin\phi_i(t), -r_0 \cos\theta_i(t))$. The parameters $r_0$, $\theta_i(t)$ and $\phi_i(t)$ are radius, polar angle, and the azimuth of the $i$th spin at time $t$ on the Bloch sphere (details in SM Section 1.1).

The dynamics of the spins, governed by the mean-field Hamiltonian associated with Eq. (1), is computationally efficient[58]. It is described by $N$ ordinary differential equations for the 2×2 density matrices $\rho_i(t)$ (SM, Section 1.1):

$$\frac{d\rho_i(t)}{dt} = -i\Omega_i[s_i^z, \rho_i(t)] - \frac{i}{2}\left(r(t)e^{-i\psi(t)}[s_i^-, \rho_i(t)] + r(t)e^{i\psi(t)}[s_i^+, \rho_i(t)]\right), \tag{3}$$

where $r$ and $\psi$ are the collective dynamical parameters that depend on the full density matrix, $re^{i\psi} = 2N^{-1}\sum_j \text{Tr}(s_j^- \rho_j)$, in analogy with the Kuramoto model[59] and the mean-field theory of superconductivity[60]. The parameter $r$ is related to the collective coherence by $\eta = r^2$. Eq. (3) governs the dynamics of the density matrix of the entire system, providing the collective coherence as a function of time.

Fig. 2 analyzes the transition between the decoherence phase and the supercoherence phase. *Decoherence phase*: In the absence of interactions ($J_{ij} = 0$), the collective coherence $\eta(t) = \eta(0)|\int p_\sigma(\Omega) e^{i\Omega t} d\Omega|^2$ decays with a rate that depends on the frequency distribution $p_\sigma$ and its variance $\sigma^2$ (black curves in the panels of Fig. 2(a)). For weak interactions or a strong frequency disorder (e.g., $\sigma = 1.5$), the collective coherence $\eta(t)$ still decays to zero (the rightmost panel of Fig. 2(a)). *Supercoherence phase*: For strong interactions or weak frequency disorder (e.g., $\sigma = 0.5$), the collective coherence $\eta(t)$ does not decay and keeps oscillating around a finite value $\bar{\eta}$ (Fig. 2(a)).

The transition between the two phases becomes sharp for $N \to \infty$, defining a critical disorder $\sigma_c$. Fig. 2(b) shows that for a disorder approaching $\sigma_c$, the oscillation period of $\eta(t)$ diverges. Fig. 2(c) shows the time-averaged coherence $\bar{\eta}$ vs. $\sigma$, identifying the phase transition from supercoherence to decoherence at $\sigma_c$. Fig. 2(d) shows the transition as a function of $\sigma$ and initial state $\theta_0$.

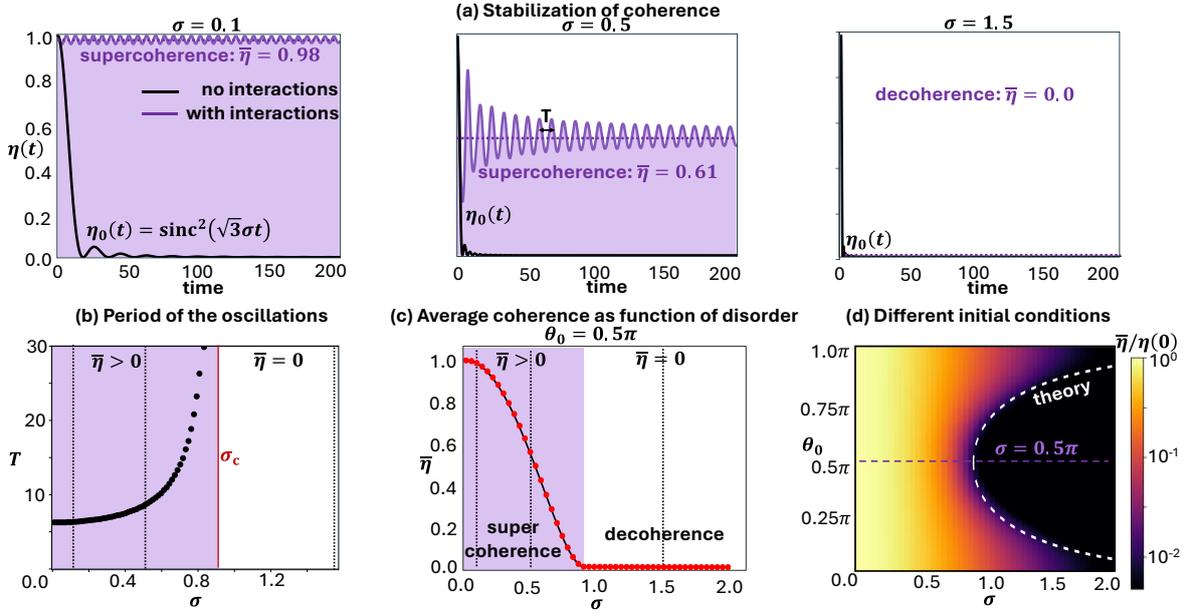

**Figure 2: Analysis of the supercoherence dynamics for all-to-all interactions $J_{ij} = -N^{-1}$. (a)** Collective coherence $\eta(t)$ for various disorder strengths $\sigma$, for an initial spin coherent state with $\theta_0 = \pi/2$. The non-interacting case ($J_{ij} = 0$) is shown as a black curve. Horizontal lines mark the average coherence $\bar{\eta}$. **(b)** The period of oscillations $T$ of the collective coherence $\eta(t)$ as a function of $\sigma$. The period of oscillations diverges when approaching the critical value $\sigma_c = \pi/(2\sqrt{3}) \approx 0.91$. **(c)** Time-averaged collective coherence $\bar{\eta}$ versus $\sigma$, for initial conditions $\theta_0 = 0.5\pi$. The parameters plotted in (a) are marked by vertical lines in (b, c). **(d)** Map of $\bar{\eta}$ versus disorder $\sigma$ and initial angle $\theta_0$. The white dashed curve shows the boundary of the supercoherence phase. The dashed horizontal line mark the values considered in (b). Theory predictions are taken from Eqs. (4), whereas numerical results are obtained solving Eq. (3) with uniform frequency distribution of $\Omega_i$ for $N = 1000$.

Our analysis in SM Section 3 shows that supercoherence survives beyond the mean-field approximation. By solving the 2$^{nd}$ order hierarchical Heisenberg equation, we find that supercoherence conserves not only classical coherence but also quantum correlations, remaining robust despite disorder (examples presented in Fig. S6). Specifically, since the interaction Hamiltonian is a spin squeezing operator[61], a global squeezing parameter (which is central in quantum sensing techniques) is preserved against disorder under the same conditions for supercoherence. In the next sections we find that the supercoherence phase also preserves higher-order correlations.

We show numerically and analytically that the supercoherence dynamical phase transition occurs for different frequency distributions $p_\sigma(\Omega)$. The critical disorder $\sigma_c$ and order parameter $\bar{\eta}$ quantifying the transition depend on the specific distribution (Fig. 3(a)) and can be found analytically for some of the cases (Table 1). The dependence of the order parameter $\bar{\eta}$ on the disorder $\sigma$ differs for uniform, Gaussian, and Lorentz distributions, yet all of them have a similar critical value $\sigma_c \sim 1$ and the same critical exponent, with $\bar{\eta} \propto (\sigma_c - \sigma)^2$ near the transition, indicating the universal character of supercoherence.

Fig. 3(b) investigates the emergence of supercoherence not only for pure states ($r_0 = 1$) but also for mixed states ($r_0 < 1$), presenting contours of critical disorder $\sigma_c$ on the Bloch ball. Interestingly in the case of low excitation (i.e., $\theta_0 \ll 1$), only the uniform distribution supports the supercoherence phase. However, in a half-excited system $\theta_0 = \pi/2$, all the frequency distributions support supercoherence. This suggests that highly excited systems are more favorable for supercoherence.

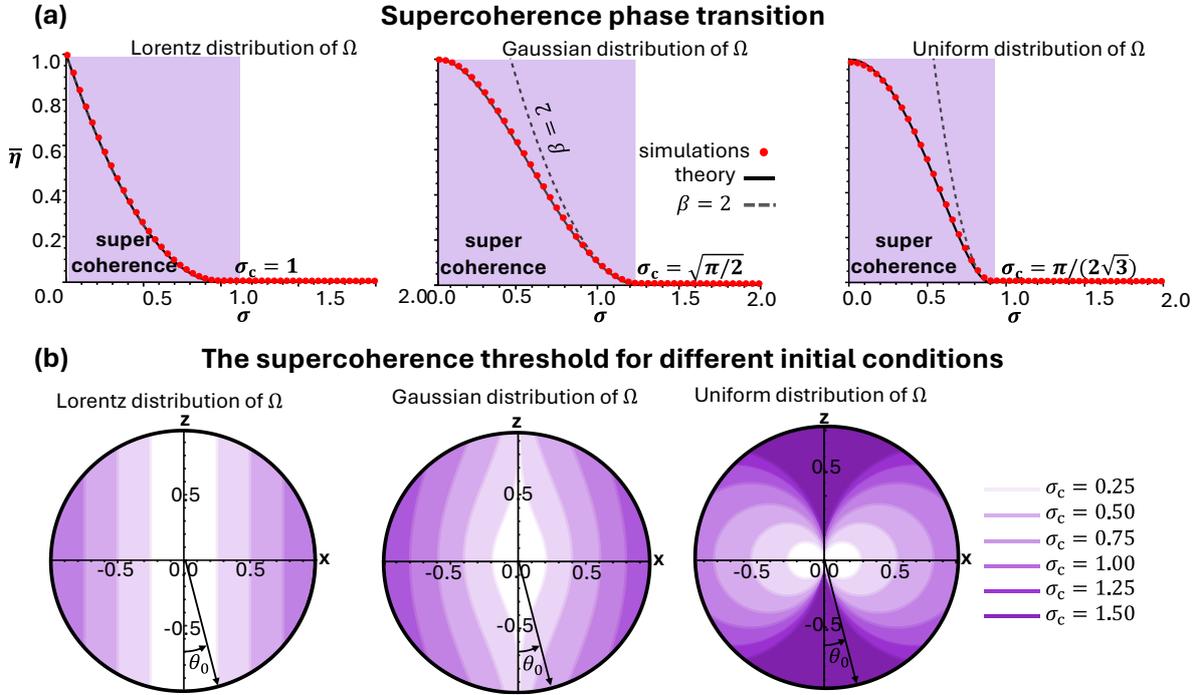

**Figure 3: The supercoherence phase transition. (a)** The order parameter $\bar{\eta}$ as a function of disorder strength $\sigma$ for various frequency distributions $p_\sigma(\Omega)$. The frequency distribution alters the exact transition point $\sigma_c$, but maintains the same critical exponent $\beta = 2$, indicating universality. **(b)** Contour plot showing the critical disorder $\sigma_c$ as a function of initial state on the Bloch ball. For $\sigma_c = 0.25$, supercoherence exists for the entire ball except the white region, for $\sigma_c = 0.5$ it exists for the entire ball except the white and light violet regions, etc. Note that uniform distribution is very different from Lorentz and Gaussian because uniform distribution is nonzero only in the finite frequency range, unlike the other distributions.

The dependence of $\bar{\eta}$ on the disorder $\sigma$ can also be found analytically by solving the following system of self-consistent equations for $r = \sqrt{\bar{\eta}}$ and for a free parameter $\Delta$ (SM, Section 1.3):

$$\begin{cases} r \int_{-\infty}^{+\infty} \dfrac{p_\sigma(\Omega) d\Omega}{(\Omega + \Delta)^2 + r^2} = \dfrac{\sin\theta_0}{r_0}, \\ \int_{-\infty}^{+\infty} \dfrac{p_\sigma(\Omega)(\Omega + \Delta)}{(\Omega + \Delta)^2 + r^2} d\Omega = \dfrac{\cos\theta_0}{r_0}. \end{cases} \quad (4)$$

The results (SM, Section 1.4) are summarized in Table 1 for various distributions $p_\sigma$.

**Table 1:** Critical disorder $\sigma_c$ and time-averaged collective coherence $\bar{\eta}$ for selected frequency distributions $p_\sigma$

| $p_\sigma$ | $\sigma_c$ | $\bar{\eta}$ ($\sigma > \sigma_c$) | $\bar{\eta}$ ($\sigma < \sigma_c$) |
|---|---|---|---|
| **Uniform distribution** | $\dfrac{\pi r_0}{2\sqrt{3} \sin\theta_0}$ | 0 | $3\sigma^2 \cot^2\left(\dfrac{\sigma}{\sigma_c}\dfrac{\pi/2}{\sin\theta_0}\right)$ |
| **Gaussian distribution** | $\sqrt{\dfrac{\pi}{2}} \dfrac{r_0}{\sin\theta_0} e^{\left(\text{erfi}^{-1}(\cot\theta_0)\right)^2}$ | 0 | numerical, see Eq. (S38) |
| **Lorentzian distribution** | $r_0 \sin\theta_0$ | 0 | $(\sigma_c - \sigma)^2$ |

To summarize the mean-field analysis, supercoherence emerges for different initial conditions, for all frequency distributions, and even for random fluctuations in the strength of interactions (SM, Sections 1.5, 1.6), which further confirms its universality.

**Supercoherence in the low-excitation regime**

So far, we analyzed the emergence of supercoherence in the case of all-to-all interactions. In this special case, supercoherence relates to previously studied dynamical phase transitions in superconductivity and cavity quantum electrodynamics[38]. We next show that supercoherence is a far broader and universal phenomenon, existing for a wide range of interaction geometries (not only for all-to-all interactions as above) and requiring only a few long-range interactions per particle. To analyze arbitrary interaction geometries, we construct a more general theoretical description that helps unravel the full quantum description of supercoherence.

The theory presented in this section brings us to the most important property of supercoherence: it preserves superpositions. That is, robustness to disorder extends beyond individual supercoherent states $|n_{\text{SC}}\rangle$ to any superposition $\sum_n c_n |n_{\text{SC}}\rangle$ with arbitrary complex coefficients $\{c_n\}$. Consequently, entangled spin states can also become inherently robust, suggesting a path to quantum memory storage.

The notoriously large Hilbert space of the many-body system of spins makes it unfeasible to solve the dynamics exactly. Fortunately, we can characterize the emergence of supercoherence already in the low-excitation regime, where all $|n_{\text{SC}}\rangle$ have number of excitations $n$ much smaller than number of spins: $n \ll N$ (i.e., $\theta_0 \ll 1$), which reduces the problem to diagonalization matrices with linear size $N$ instead of $2^N$. The price to pay is that the low-excitation regime lacks a phase transition when increasing $\sigma$. Thus, only by combining the results from the previous section with the results of this approach—we can fully capture all the properties of supercoherence.

We exploit the Holstein-Primakoff approximation[62], which is valid for weak excitations ($\theta_0 \ll 1$), enabling to change spin operators to boson operators[54]: $s_i^z \to 1/2 - a_i^\dagger a_i$, $s_i^- \to a_i^\dagger$. Then the Hamiltonian in Eq. (1), up to the constant, can be approximated as (SM, Section 4.1):

$$H \approx \sum_i \Omega_i a_i^\dagger a_i + \sum_{i,j} J_{ij} a_i^\dagger a_j. \quad (5)$$

Being quadratic, $H$ can be solved, for arbitrary frequencies $\Omega_i$ and interactions $J_{ij}$, by diagonalizing an $N \times N$ matrix $\delta_{ij}\Omega_i + J_{ij}$. Interestingly, Eq. (5) also describes a wide range of photonic systems, thus showing that supercoherence can also be found in such platforms, and is not limited to spin systems.

**Supercoherence in synthetic systems of different interaction geometries**

The interactions $J_{ij}$ allow to realize synthetic systems with different geometries supporting supercoherence. Mapping each spin to a network vertex and each nonzero interaction $J_{ij} \neq 0$ to an edge, our system can be represented as a network. Table 2 summarizes four network types: regular lattices, small-world networks[63], Erdos-Renyi networks[64], and Barabasi-Albert networks[65]. The first column of Table 2 shows the lattices, parametrized by the dimensionality, i.e. 1D, 2D, 3D, etc. The second column shows small-world networks[63], characterized by a rewiring probability $p$ and number of nearest-neighbor edges $k$. Constructing such networks begins with a 1D circular geometry, and subsequently rewiring one end of each edge randomly with probability $p$. This procedure allows for tuning the network from short-range interactions at $p = 0$ to long-range interactions at higher $p$ values, with the extreme case of a random network at $p = 1$. This process maintains fixed the total number of interaction edges, $Nk/2$. The third column shows Erdos-Renyi networks[64], characterized by the probability $p$ to have a link between any two vertices. Such networks always have long-range connectivity for any $p > 0$. The fourth column shows the Barabasi-Albert model[65], where the parameter $m$ determines the number of edges that each new vertex creates when joining the network, connecting to existing vertices preferentially based on their degree.

At the bottom row of Table 2, we simulate, for each type of network, the time-averaged relative coherence $\bar{\eta} = \eta(t)/\eta(0)$, the connectivity, and the energy gap separating the supercoherent state $|n_{SC}\rangle$ from the continuum of energy states (detailed in the next section). The connectivity is defined as the average number of connections for each vertex divided by $N - 1$, yielding 1 for all-to-all interactions and $\sim N^{-1}$ for regular lattices.

In regular lattices (left-most column) the coherence decays to zero for any $\sigma$ ($\bar{\eta} = 0$), indicating lack of supercoherence for any finite dimension. By contrast, in networks with long-range interactions (all other columns) the coherence $\bar{\eta}$ is preserved for high enough $p$. Small-world networks indicate that supercoherence can exist even with very low connectivity, with $k = 4$ connections per spin being already sufficient. Barabasi-Albert networks can also support supercoherence for any $m > 1$. Such networks are ubiquitous in many areas of science, showing the prospects of finding supercoherence in various systems of interest.

**Table 2:** Various synthetic quantum systems can support supercoherence (highlighted by pink glow). All simulations assume a fixed disorder $\sigma = 0.2$ and $N = 1000$.

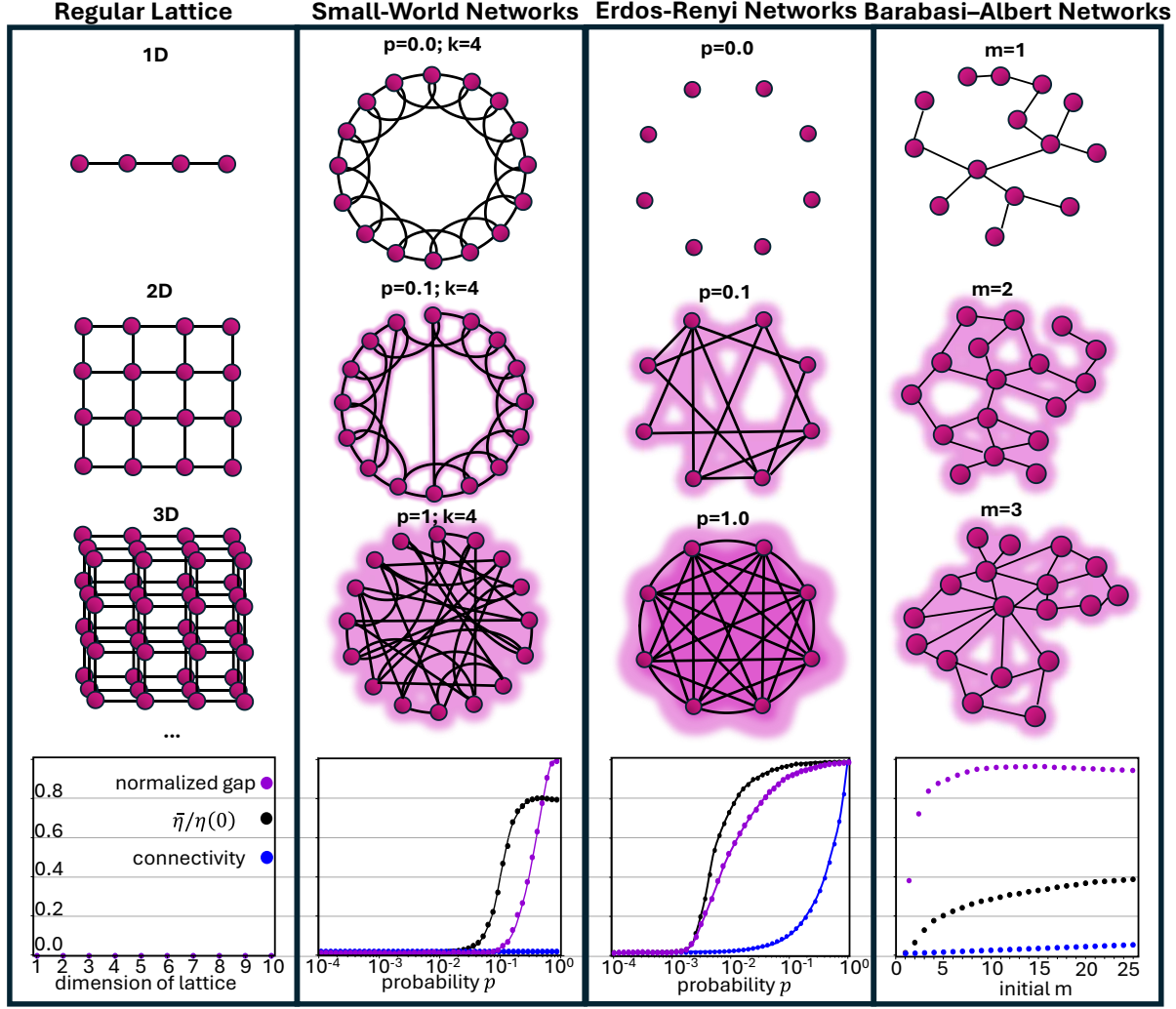

**The supercoherence energy gap and isolated states**

In this section, we show that supercoherence is linked to isolated, delocalized energy states separated by sizeable energy gaps from continuous bands of other energy states. We calculate the energy spectrum by diagonalizing Eq. (5), yielding $N$ non-interacting harmonic oscillators $H = \sum_k E_k b_k^\dagger b_k$. The eigenvalues $E_k$ of the matrix $\Omega_i \delta_{ij} + J_{ij}$ correspond to new normal mode operators $b_k^\dagger = \sum_i a_i^\dagger v_{ki}$, with $v_{ki}$ the $i^{\text{th}}$ element of the $k^{\text{th}}$ eigenvector.

Fig. 4(a) shows that supercoherence exists in the low-excitation regime since the coherence does not decay with time. Supercoherence occurs when a gap emerges, separating one eigenvector from the band formed by all the others. This eigenvector is a supercoherent state $|n_{\text{SC}}\rangle$. Fig. 4(b) shows the formation of the gap, which decreases when the disorder strength $\sigma$ is increased. The non-zero time-averaged coherence and the gap are directly connected as shown in Fig. 4(c). Fig. 4(d) shows that such a gap and an isolated state $|n_{\text{SC}}\rangle$ both emerge for each excitation number $n$, forming a harmonic oscillator of supercoherent states with an equal ladder spacing of $\langle \omega \rangle = N^{-1} \sum_i \omega_i$ between consecutive states. We also conducted numerical simulations of the full system (with $n_{\text{SC}}$ up to 3), demonstrating close agreement with the Holstein-Primakoff approximation (SM Section 4.5). The direct connection between the coherence preservation and the gap formation (SM Section 4.3) is shown in Table 2: every network with non-zero coherence (violet dots) also has gaps (black dots). For the case of all-to-all interactions, we prove analytically the existence of an energy gap $E_{\text{gap}} = \sqrt{3}\sigma(\coth(\sqrt{3}\sigma) - 1)$ and average fidelity $\bar{\eta}/\eta(0) = 9\sigma^4 \operatorname{csch}^4(\sqrt{3}\sigma)$ (SM Sections 4.2 and 4.4), accurately matching the numerical simulations presented in Fig. 4(c).

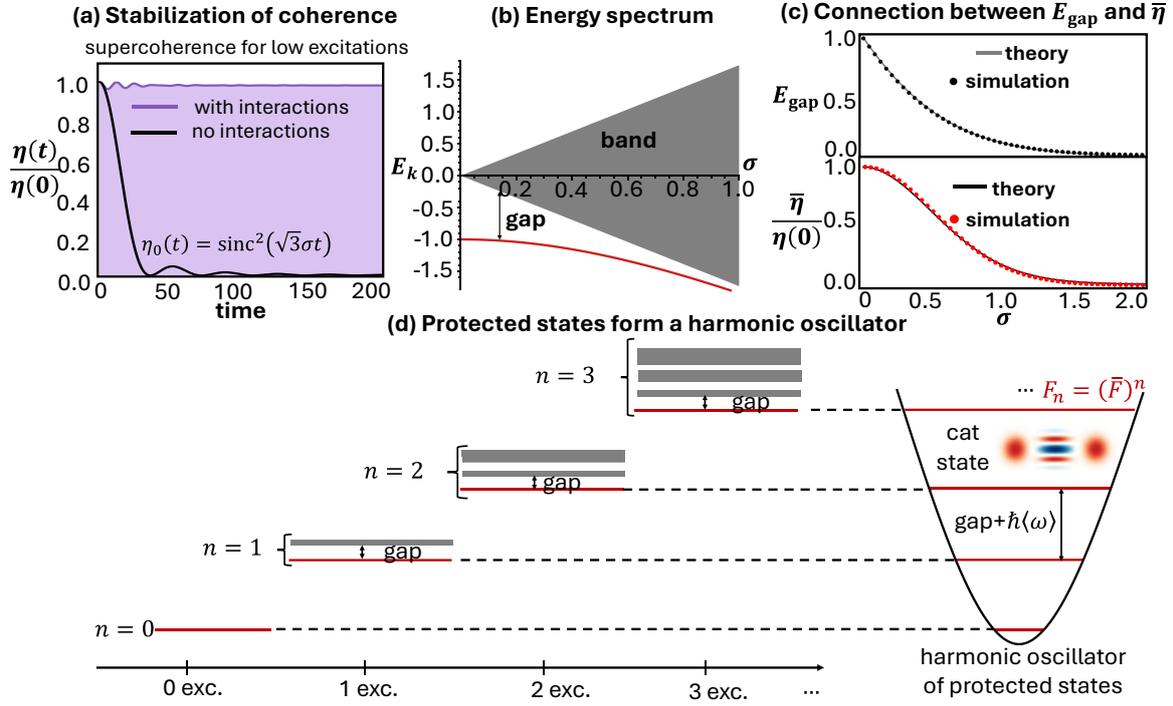

**Figure 4: The supercoherence energy gap and quantum memory: results in the low-excitation regime.** (a) Supercoherence exists also in the low-excitation regime, as seen in the dynamics of the relative coherence $\eta(t)/\eta(0)$ for $\sigma = 0.1$ and $\theta_0 \ll 1$. (b) Energy spectrum $E_k$ of the normal modes $b_k^\dagger$ after diagonalization, showing an isolated state and a gap that decreases for higher disorder $\sigma$. (c) $E_{\text{gap}}$ and $\bar{\eta}/\eta(0)$ versus $\sigma$, showing that a larger gap directly relates to coherence of higher stability. This relation enables using the gap as a signature of supercoherence. (d) As highlighted by the Holstein-Primakoff approach, the total spectrum is approximated by the direct sum of single-excitation spectra, forming a harmonic oscillator of isolated protected states $|n_{\text{SC}}\rangle$. Thus, we can preserve the coherence against disorder any superposition of harmonic oscillator Fock states $\sum_n c_n |n_{\text{SC}}\rangle$ (e.g., the illustrated cat state). Simulations in this figure use a uniform frequency distribution $[-\sqrt{3}\,\sigma, +\sqrt{3}\,\sigma]$ and all-to-all interactions $J_{ij} = -N^{-1}$. Note that for $J_{ij} = +N^{-1}$ the spectrum would be inverted, i.e., each isolated state will appear above (instead of below) its corresponding energy band. However, the gap and the dynamics of the coherence would be the same. The resulting isolated delocalized supercoherence states (red lines at the bottom of each excitation $n$) and harmonic oscillator behavior are general and apply to different interaction geometries. The numerical results are obtained for $N = 1000$.

The regime of low-excitation states ($\theta_0 \ll 1$) received attention in the literature in the context of a generalized "cavity protection" mechanism[52–55], but so far only for all-to-all interactions. We find that supercoherence strongly depends on the frequency distributions: Fig. 3(b) shows that the supercoherence phase (violet) does not occur near the bottom of the Bloch sphere for the Lorentz and Gaussian distributions, which are limited to the decoherence phase (white).

In contrast, for the uniform distribution and generally for any finite-frequency distribution, Fig. 3b shows that the supercoherence phase (violet) emerges at low excitations. Nevertheless, our findings outside the low-excitation regime (e.g., Figs. 2-3 and Table 1) show that supercoherence and its various properties appear at high-excitation regime for any probability distribution. These findings suggest that conclusions drawn for uniform distribution are applicable for other distributions once going to higher excitations ($\theta_0 \sim 1$). Consequently, we use the uniform frequency distribution for simulations in Fig. 4 and Table 2.

**Quantum memory in the supercoherent phase**

In this section, we show that supercoherence can preserve *all* the quantum properties of the initial state and enables the storage of quantum states despite disorder. This analysis goes beyond the preservation of the classical coherence and second-order correlations that we showed above. Within the low-excitation regime, the isolated supercoherent states form a harmonic oscillator (Fig. 4(d)), with the $n^{\text{th}}$ isolated state $|n_{SC}\rangle$ approximating the symmetric state with $n$ excitations. A general superposition of supercoherent states $|\Psi(t)\rangle = \sum_n c_n(t)|n_{SC}\rangle$ evolves in time with $c_n(t) = c_n(0)\exp(-in\langle\omega\rangle t)$, such that the effect of disorder is mitigated. The frequency distribution alters the dynamics only through its mean frequency $\langle\omega\rangle$ that determines the separation between the isolated supercoherent states.

An even more important finding is that the time-averaged fidelity $\bar{F}_n = \lim_{T\to\infty} T^{-1}\int_0^T |\langle n_s|\Psi(t)\rangle|^2\, dt$ between the evolved state $|\Psi(t)\rangle$ and the initial symmetric state $|n_s\rangle$ is preserved in the supercoherent phase, and equals (SM Sections 4.3-4.5):

$$\bar{F}_n \approx \left(\bar{\eta}/\eta(0)\right)^n, \tag{6}$$

using the relative coherence $\bar{\eta}/\eta(0)$ described in the previous section. Eq. (6) holds for any general interaction geometry $J_{ij}$. This closed-form result matches the numerical simulations, quantifying the efficiency of the preservation of quantum states, akin to conceptual ideas in atomic and solid-state systems[53,54].

**Discussion and outlook**

Radiative decay and other decoherence mechanisms will inevitably limit supercoherence and were neglected in Eq. (1). We test their influence in SM Section 2, finding that supercoherence persists beyond the timescales of disorder-induced decoherence, but later decays on the longer timescales of radiative decay and external decoherence. Several strategies can be employed to mitigate these decoherence channels. For example, radiative decay can be suppressed by selecting transitions that are considered forbidden by selection rules or by engineering the electromagnetic environment to modify the local density of states, using photonic crystals, cavities, or metamaterials[66]. These approaches could further extend the lifetime of supercoherent states, allowing for more robust observation and utilization of this phenomenon.

The concept of supercoherence, i.e. $\bar{\eta} > 0$, is applicable more broadly than analyzed here. The theoretical model we considered in Eq. (1) describes two-level systems. However, as we show in Eq. (5), supercoherence can also arise for coupled harmonic oscillators. In another example, can occur in systems with a continuum of energies, such as electrons in solids or free particles in plasma or liquid.

Supercoherence can also occur in various synthetic systems, such as cavity systems[31,52–55] and circuit QED[37,67], where related effects have been investigated for extending coherence times and observing dynamical phase transitions. Recent progress in tweezer arrays[25] and waveguide QED[27], enable creating synthetic interaction networks with suitable parameters. Other platforms, including ion traps[33], also offer precise control over interaction strengths and selective long-range coupling, making them ideal for exploring supercoherence.

To explore novel ideas that can emerge from further investigation of supercoherence, it is interesting to highlight the relation of supercoherence to topological insulators[68]. While topological insulators rely on symmetry-protected edge states, supercoherence generates delocalized, gap-separated bulk states, preserving robust quantum properties. Between those distinct ideas, there could exist other phases and dynamical phase transitions with topological properties that depend on complex network geometries.

Another possible domain are biological systems[69], despite the current belief that most quantum effects are highly improbable, primarily due to the rapid decay of collective coherence in disordered environments. Interestingly, certain network geometries that are naturally occurring in biological systems resemble the ones that were shown here to support collective coherence (classical or quantum) through supercoherence. This is for example the case with the Barabasi-Albert model[65] and its variants, often used to model the scale-free networks that are ubiquitous in biological systems. Famous examples include protein-protein interaction networks[70], neural networks[71], as well as molecular and polymer networks[72]. The relevance of this work to biological systems implies that, although speculative, future insights could be gained from analyzing specific network geometries that appear in biological systems. Such networks could be tested for their suitability to support supercoherence, its analogue, or approximations—suggesting that it might be more widespread in nature than expected.